\documentstyle[12pt]{article}
\begin{document}

\begin{center}
{\large \bf Decimation and survival at baryon violating gantlet}

\vspace{1cm}

{\large \bf SUBIR MOHAN }

\vspace{.8cm}

{\large \bf Abstract}
\end{center}

\vspace{.6cm}

We find that for second and weakly first order electroweak phase
transition (EWPT) the mere presence of non-zero Majorana masses for
left-handed neutrinos is sufficient to ensure the destruction of any
existing baryon (lepton) asymmetry. Even if the EWPT is strongly first
order, a baryon asymmetry generated before EWPT is seen to only barely
survive, to the present, for cosmologically interesting values of neutrino
masses and mixing angles; the scenario for survival being particularly
bleak in the presence of an $SU(2)_R$ gauge symmetry at intermediate
scales. Two sets of models, presented by us earlier, that can avert the
destruction of baryon asymmetry for any value of neutrino masses and
mixing angles and any order of EWPT are briefly discussed and their
relevance, in the light of latest observations, is pointed out.

\pagebreak

The neutrinos are massless in the Standard Model, but there exist
observational indications that they may, afterall, be massive. The solar
electron-neutrino $(\nu_e)$ and the atmospheric muon-neutrino
$(\nu_{\mu})$ deficits, and the need for some hot dark matter are all
looked upon as evidence for the existence of neutrino masses [1].

It is widely accepted that the solar $\nu_e$  deficit can be explained by
the oscillation of $\nu_e$ to other neutrino species : the well known MSW
effect [2]. The atmospheric $\nu_{\mu}$ deficit is most likely to be due to
the oscillation of $\nu_{\mu}$ to $\nu_{\tau}$ [3]. The large amount of
data available on the extent of structure in the universe on a wide range
of distance scales is best fit by the Cold Hot Dark Matter (CHDM) models,
and the most successful of these models requires the hot dark matter to
account for $20 \%$ of the energy density of the universe [4]. For a flat
universe and a value of the Hubble constant (H) around $50$ Km. $s^{-1}$.
$Mpc^{-1}$, this puts a bound on the sum of stable neutrino masses $\sum
m_{{\nu}_i} \sim 5eV$.
The neutrino oscillations do not constrain the neutrino masses directly
but they impose restrictions on the mass squared difference, $\Delta
m_{ij}^2 =  |m_i^2 - m_j^2|$, and the mixing angles $\theta_{ij}$.

Small neutrino masses (in the eV range or smaller) can be naturally
generated by the see-saw mechanism [5]. All the left-handed neutrinos
$\nu_i$ in the Standard Model can have right-handed companions $N_i$ that
are Standard Model gauge group, $SU(3)_C \times SU(2)_L \times U(1)_Y$,
singlets. The $N_i$ can have Majorana mass terms of the form
$M_i(\overline{N_i^C}N_i +h.c.)$, where C denotes charge conjugation. The mass
$M_i$ can be much larger than the electroweak(EW) symmetry breaking scale,
as it may arise when a right-handed symmetry is broken or it may just be
present explicitly. When the EW symmetry is broken, Dirac mass terms of
the type $m_{d_i}(\bar{\nu_i} N_i+ h.c.)$ can arise. In the presence of
both Majorana and Dirac masses, the mass eigenvalues are $O({m_{d_i}^2}/M_i)$
and $O(M_i)$ corresponding to the self-conjugate mass eigenstates $(\nu_i^C
+ \nu_i) \equiv \omega_i$ and $(N_i^C + N_i) \equiv \chi_i$, respectively.${}^{\#1}$
Thus, a massive neutrino is essentially two self-conjugate Weyl states
with distinct masses. Typically, $m_{d_i} $ is expected to be of the same
order as the mass of the charged lepton and/or quarks of the $i^{th}$
generation. And if $m_{d_i} << M_i$, then we have a light Majorana
neutrino $\omega_i$.

The Majorana mass terms for the neutrinos violate the lepton number $(L)$
symmetry. At temperatures well above the electroweak phase transition
(EWPT) temperature $(T_{EW})$ scale, lepton number may be violated due to
the presence of Majorana mass for right-handed neutrinos $(N_i)$. Two
lepton number violating processes are:


(i) decay of massive right-handed neutrinos $N \rightarrow l_L \phi, l_L^C
\phi^C$, where $l_L$ is a left-handed lepton doublet and $\phi$ is the
standard electroweak Higgs doublet, and

(ii) $N$ mediated $2 \leftrightarrow 2 $ scatterings with an effective
interaction term $(l_L  l_L \phi \phi +h.c.)$. 


\noindent If these processes are in thermal equilibrium at the same time as the
sphaleron interactions, then the baryon number $(B)$ and $L$ asymmetries
will be wiped out even if initially $(B-L) \neq 0$.

The requirement that the rate of the lepton number violating interactions
$(\Gamma_{\Delta L \neq 0})$ should be less than the expansion rate of
the universe $(H)$, so that these interactions are not in thermal
equilibrium simultaneously with the sphaleron interactions, at ${10}^{12}
GeV >T>T_{EW}$,
has been extensively used to put restrictions on the values of Majorana
masses $(m_{\nu_i})$ of the left-handed neutrinos [6]. The upper limits
obtained range from $10^{-3} eV$ to $10^5 eV$ depending on the specific
lepton number violating process considered and other details of the
mechanisms that may be effective in protecting the baryon number
asymmetry. It should be noted that even though the analyses carried out at
$T>T_{EW}$ can constrain the Majorana masses of the left-handed neutrinos,
the left-handed neutrinos can acquire a mass only after the electroweak
symmetry breaking at $T \sim T_{EW}$.

We find that if the sphaleron interactions are in thermal equilibrium
after EWPT and if the left and the right handed neutrinos have non-zero
Majorana masses, then $B$ and $L$ will be driven to zero {\it regardless
of the values of the Majorana masses}.

And it appears that the sphaleron interactions may well be in thermal
equilibrium after EWPT, at least in the models with just one Higgs doublet.

The sphaleron interactions will not be in thermal equilibrium below
$T_{EW}$ if the EWPT is sufficiently strongly first-order such that
${E_{sph}}/T_{EW} {^> _{\! \sim}} 45 $ where $E_{sph}$ is the energy of
the sphaleron configuration [7]. This constrains the mass of the neutral
Higgs boson $(m_H)$ to be $m_H {^< _{\! \sim}} (35 -80)$ GeV [8]. The
experimental lower limit $m_H > 58 GeV$ may be consistent with a strong
first-order EWPT. But the one-loop quantum corrections to the electroweak
scalar potential due to the heavy top quark $(m_{\rm top} \sim 174 GeV)$
will render the zero temperature effective potential for one Higgs doublet
unbounded from below unless $m_H > 130 GeV$ [9]. Thus, in the one Higgs
doublet models if the vacuum is to be stable then EWPT {\it cannot} be
strongly first-order. In the presence of more than one Higgs doublet the
constraints are different [10] but, still, there is no reason for the scalar
potential parameters to forbid a second order or weakly first-order EWPT.

For a strongly first-order EWPT a non-zero $B$ may be generated through
electroweak baryogenesis [11]. However, we should note that this mechanism is
not very well understood and a debate on its viability is still on [12]. A
baryon asymmetry, with $(B-L) \neq 0$, generated well above $T_{EW}$ may
survive the combined onslaught of sphaleron and $B$ and/or $L$ violating
interactions, but only barely for cosmologically interesting values of
neutrino masses and mixing angles, as we shall show. And thus, the need
for viable and well understood mechanisms and models that can protect the
baryon asymmetry generated well above $T_{EW}$ and/or generate it after
the sphaleron interactions have gone out of equilibrium cannot be overstated.

First we shall perform an exercise in equilibrium thermodynamics along the
lines of Harvey and Turner [13] to show that the presence of Majorana mass
for neutrinos is sufficient to completely erase any baryon asymmetry if
the sphaleron interactions are in thermal equilibrium after EWPT. Then we
demonstrate that for the cosmologically interesting values of neutrino
masses and mixing angles survival of a baryon asymmetry generated well
above $T_{EW}$ is only barely possible even if the sphaleron interactions
are not in thermal equilibrium after EWPT. And finally, we briefly discuss
and stress the importance of two models, that we had constructed earlier
[14,15], that allow the baryon asymmetry produced by the decay of heavy
GUT scalars to be the baryon asymmetry that is observed today for any value
of neutrino masses.

{\bf 1.} Particle asymmetries are most conveniently expressed in terms of
chemical potentials. For ultrarelativistic particles the relation between
the excess of particle over antiparticle and the particle's chemical
potential is given by [13,16]

$$n_+ -n_-  =  \frac{gT^3}{3} \left(\frac{\mu}{T}\right) F_b(\frac{m}{T}) \quad
{\rm (bosons)}, \    \eqno(1a) $$

$$n_+ -n_-  =  \frac{gT^3}{6} \left(\frac{\mu}{T}\right) F_f(\frac{m}{T}) \quad
{\rm(fermions)}, \    \eqno(1b) $$

$$F_b(x)  =  \frac{3}{\pi^2} \int_x^\infty dy \; y \sqrt {y^2 - x^2} \frac
{e^y}{{(e^y-1)}^2} \ , \  \eqno(1c) $$

$$F_f(x)  =  \frac {6}{\pi^2} \int_x^\infty dy \; y \sqrt{y^2-x^2}
\frac{e^y}{{(e^y+1)}^2}\ , \ \eqno(1d)$$

\noindent where $n_+(n_-)$ is the equilibrium number density of the particle (CP
conjugate) species and $\mu(m)$ is its chemical potential (mass) while $g$
counts the internal degrees of freedom. We have assumed that $|\mu/T|, \
m/T << 1$.

For $M_W << T< T_{EW}$ the particles expected to be in chemical equilibrium
are $N$ standard model generations of fermions, the components of m
Standard Higgs doublets that have not been {\it eaten up} by $W^{\pm}$ and
$Z$, and the usual gauge bosons of $SU(3)_C \times SU(2)_L \times U(1)_Y
$. Rapid electroweak interactions enforce the following equilibrium
relations among the chemical potentials: 

$$ \mu_W= \mu_- + \mu_0  \quad    (W^- \leftrightarrow \phi^- +\phi^0),
\ \eqno(2a)$$

$$\mu_{dL}= \mu_{uL} + \mu_W  \quad    (d_L \leftrightarrow u_L+W^-),
\ \eqno(2b)$$

$$\mu_{iL}= \mu_i+ \mu_W  \quad    (i_L \leftrightarrow \nu_{iL} + W^-),
\ \eqno(2c)$$

$$\mu_{uR}= \mu_0 + \mu_{uL}  \quad    (u_R \leftrightarrow \phi^0 + u_L),
\ \eqno(2d)$$

$$\mu_{dR}= - \mu_0 + \mu_{dL}  \quad    (d_R \leftrightarrow {\bar \phi^0} +d_L),
\ \eqno(2e)$$

$$\mu_{iR}= - \mu_0 + \mu_{iL}  \qquad    (i_R \leftrightarrow {\bar \phi^0} +i_L).
\  \eqno(2f)$$

\noindent In our notation the relationship between chemical potentials and the
particles in brackets is one to one. $i$ denotes a lepton species $(e^-,
\mu^-, \tau^-)$. Cabibbo mixing should maintain the equality of chemical
potentials of the up and down-quark states of different generations, and
we assume that mixing between the components of $m$ Higgs doublets
maintains the equality of their chemical potentials.

So long as sphaleron
interactions are rapid, the following relation among the chemical
potentials is enforced :

$$N(\mu_{uL} + 2 \mu_{dL}) + \sum \limits_i \mu_i = 0. \eqno(3)$$

The charge, baryon and lepton numbers carried by particles in chemical
equilibrium can be expressed in terms of the chemical potentials as:

$$B  =  N(\mu_{uL} + \mu_{uR}) + N(\mu_{dL} + \mu_{dR})  =   4N \mu_{uL}
+2 N \mu_W , \ \eqno(4)  $$

$$L  =  \sum\limits_i (\mu_i+\mu_{iL} + \mu_{iR}),  \    \eqno(5)$$
$$Q  =  2N(\mu_{\mu L} + \mu_{uR}) - N(\mu_{dL} + \mu_{dR}) -\sum\limits_i 
(\mu_{iL} +\mu_{iR}) -6 \mu_W -2(m-1) \mu_{-}. \  \eqno(6)$$
 
The eight gluon fields and $Z$ and photon fields have vanishing chemical
potential and have been ignored for this exercise. Because of the vacuum
condensate of $\phi^0$ Higgs bosons, $\mu_0$ must be equal to zero. And
since the Dirac mass terms for fermions mix the left and right-handed
states, their chemical potentials must be equal 

$$ \mu_{uL} = \mu_{uR}, \quad \mu_{dL} = \mu_{dR}, \quad \mu_{iL} = \mu_{iR}.  \eqno(7)$$

The Majorana mass terms for the neutrinos mix the neutrinos and
antineutrinos thereby making their chemical potential zero,

$$ \mu_i =0.   \eqno(8) $$

Using (2), (3), (7) and (8) we have

$$ Q= \left(\frac{-10}{3} N -12 -2(m-1)\right) \mu_W, \eqno(9)$$

\noindent and if the mass effects are also taken into account then 

$$Q=\left[\frac{-8}{3} \sum\limits_u F_f (x_u) - \frac{2}{3} \sum\limits_d F_f
(x_d) - 2\sum\limits_i F_f (x_i) - 6 F_b (x_W) - \right.$$

$$ \left. 2(m-1) F_b(x_{\phi^-})\right] \mu_W, \eqno(10) $$

\noindent where $x_j \equiv m_j/T$.

The electric charge carried by the particles in chemical equilibrium must
be zero unless some special strategy has been adopted to ensure a non-zero
value. For $Q=0$, clearly $\mu_W =0$; and all other chemical potentials
are also zero. Thus, $B=0(=L)$. This is our main point. 

There is another way of easily seeing the physical basis of this result. 
In the presence of
Majorana mass, the left-handed (and also right-handed) neutrinos cannot be assigned a definite
lepton number. And consequently the charged weak interactions $(i_L
\leftrightarrow \nu_{iL}+W^-)$ rapidly violate lepton number; even the
sphaleron interactions can no longer preserve $B-L$. Hence, $B(L)$ is
driven to zero.

While the sphaleron
interactions are in thermal equilibrium for $M_W << T< T_{EW}$, the
top-quark and the Higgs scalars (both neutral and charged) may become
non-relativistic and disappear from the thermal soup due to annihilations
and decay but $B$ will remain zero.

{\bf 2.} If the EWPT is strongly first order, the observed baryon
asymmetry could have been produced by: 

(i) the out-of-equilibrium decay of heavy GUT (type) scalars,$^{\#2} $

(ii) the baryogenesis via leptogenesis mechanism, involving the out of
equilibrium decay of heavy right-handed Majorana neutrinos and the
subsequent conversion of the lepton asymmetry thus generated into a
comparable baryon asymmetry by the sphaleron interactions [18], and

(iii) the electroweak (EW) baryogenesis mechanism [11].

We have noted, earlier, that the EW baryogenesis mechanism is not well
understood and its viability is, still, a contentious issue [12]. It is, more
or less, ruled out in the one Higgs doublet model while with two Higgs
doublets it may generate an adequately large baryon asymmetry, but only 
marginally [19].

Let us see how the cosmologically interesting values of neutrino masses
and mixing angles affect the survival of baryon asymmetry generated by the
out-of-equilibrium decay of heavy GUT (type) scalars and by the
baryogenesis via leptogenesis mechanism.

If the solar $\nu_e$ and the atmospheric $\nu_{\mu}$ deficits can be
accounted for by the $\nu_e \rightarrow \nu_{\mu}(\nu_{\tau})$ and
$\nu_{\mu} \rightarrow \nu_{\tau}$ oscillations respectively, then the
constraint on the sum of the left-handed Majorana neutrino masses due to
dark matter requirements, $\sum\limits_i m_{\nu_i} \sim 5$ eV, requires
$m_{\nu_e}\sim m_{\nu_{\mu}}\sim m_{\nu_{\tau}}\sim 1.6$ eV [1]. And
$sin^2 2\theta_{ei} \sim (0.4-1.5) \times 10^{-2}, \ sin^2 2 \theta_{\mu
\tau} \sim 1.0$.

A somewhat more successful model for the large scale structure formation
in the universe [4] requires $m_{\nu_{\mu}} \sim m_{\nu_{\tau}} \sim 2.4$ eV,
 $sin^2 2 \theta_{\mu \tau} \sim 1.0$. In this case the solar $\nu_e$
deficit is accounted for by the oscillations of $\nu_e$ into a sterile
neutrino species $(\nu_s)$. Both $\nu_e$ and $\nu_s$ are required to be lighter than
$2.4$ eV, though the exact value of their masses is not determined. Even
if $\nu_e$ is lighter than $\nu_{\mu}$ and $\nu_{\tau}$, we think it is
safe to assume that the $e- \mu$ flavor mixing angle is not very small. It
should be reasonable to suppose that it is of the same order as the mixing
angle for the first and second generation quarks, $sin \theta_{e \mu} \sim
sin \theta_{u s} \sim 10^{-2} - {10}^{-1}$.

Broadly, the cosmologically interesting values of the lepton flavor mixing
angles can be taken to be of the same order as the quark mixing angles:
$sin \theta_{e \mu} \sim sin \theta_{\mu \tau} \sim {10}^{-2} - 10^{-1}$; and two of the left-handed neutrinos $(\nu_{\mu}, \;
\nu_{\tau})$ should have mass in the $1-5$ eV range (with $\Delta m_{\mu
\tau}$ small) while $\nu_e$ can be much lighter.

For $m_{\nu_i}$ in the range $1-5$ eV, the lepton number violating
$2 \leftrightarrow 2$ scatterings mediated by the heavy right-handed
Majorana neutrinos will be in thermal equilibrium for
$(10^{10}-10^{11}) GeV  {^<_{\! \sim}} T< M_i$.$^{\#3}$ If $sin^2 \theta_{e
\mu}$ and $sin^2 \theta_{\mu \tau}$ are larger than $10^{-5}-10^{-4}$,
then the lepton flavor mixing due to charged weak (W mediated)
interactions will also be in thermal equilibrium at $T \sim
(10^{10}-10^{11})$ GeV and all the lepton numbers will be violated.

If any of the right-handed neutrino Majorana masses $(M_i)$ are smaller
than $10^{10}$ GeV and the corresponding $m_{\nu_i}$ is larger than
$10^{-3}$ eV, then for some temperature range around and above $M_i$ the
lepton number violating decay of the right-handed neutrino will be in
thermal equilibrium.$^{\#4}$ And at $M_i < 10^{10} $ GeV the value of lepton
flavor mixing angles $sin^2 \theta_{ij}$ need only be larger than $(M_i
GeV/10^{15} GeV)$ for $W$ mediated lepton flavor mixing to be in thermal
equilibrium: a constraint that should be easily satisfied for
cosmologically interesting cases.

We, thus, see that the baryon asymmetry produced at $T {^>_{\! \sim}}
10^{10}$ GeV has to run through the gantlet of lepton number violating
$2\leftrightarrow 2$ scatterings and decays of heavy right-handed Majorana
neutrinos. These lepton number violating processes alongwith rapid lepton
flavor mixing and sphaleron interactions can potentially destroy $B$ even
if $(B-L) \neq 0$. For $B$ produced at $T< 10^{10} $ GeV, with $(B-L)
\neq 0$, the dangerous processes are the in-equilibrium decays of
right-handed neutrinos in case some of the $M_i$'s are less than $10^{10}$
GeV.

We, now, consider the survival of baryon asymmetry, generated well above
$T_{EW}$, in the absence  (and, then in the presence) of an intermediate
$SU(2)_R$ guage symmetry.

\vspace{.4cm}

{\bf 2.a.} It is well known that without an $SU(2)_R$ gauge symmetry
somewhere in the region $T_{EW}< T <T_{GUT}$, a unification of the gauge
coupling constants is not possible [20] without invoking the existence of
split multiplets of fermions [21,22] (or supersymmetry [21]). Still, heavy
$GUT$ (type) scalars can exist and, surprisingly, the baryon asymmetry
generated in their out-of-equilibrium decay has a better chance of
surviving than when an $SU(2)_R$ gauge symmetry is present (as will be
seen in sec.2.b.).

The heavy $GUT$ scalars present in the primeval soup must decay well
before the temperature falls below $10^{10}$ GeV if they have regular
strength Yukawa couplings with the fermions $(f_{\rm top} \sim 1, \ f_{\rm
up} \sim 10^{-5} - 10 ^{-6})$. This is because the $GUT$ scalars with
regular strength Yukawa coupling with the up and down quarks cannot be
lighter than $(10^{10} - 10^{11})$ GeV without reducing the proton
lifetime ($t_P$) below the current experimental lower limit, $t_P {^>_{\!
\sim}} {5.10}^{32}$ years [23]; and their couplings to heavier fermions
ensure that their decays are in thermal equilibrium for $T>{10}^{10}$ GeV.

The baryon asymmetry generated at $T>{10}^{10}$ GeV faces the double risk
of being decimated by the lepton number violating scatterings and the
in-equilibrium decays of heavy right-handed Majorana neutrinos.

However, since the right-handed electrons $(e_R)$ enter chemical
equilibrium at $T {^<_{\! \sim}} {10}^4 $ GeV it has been suggested that
they may act as repositories of lepton number which is transformed into a
comparable baryon number by the sphaleron interactions at $T< {10}^4$ GeV
[24] provided there are no lepton number violating processes below ${10}^4$
GeV ($M_i$'s $> {10}^4$ GeV, or if an \linebreak $M_i <{10}^4 $ GeV then $m_{\nu_i} <
{10}^{-3} $ eV). But producing an adequately large number of $e_R$'s in
scalar decays is by no means easy. The ratio of partial decay rates of a
heavy GUT (type) scalar into two distinct modes is roughly proportional
to the square of the smaller of ($f_i^U/f_j^U, \ f_i^D/f_j^D$), where
$f_k^U(f_k^D)$ is the Yukawa coupling for the up(down) sector fermions in
the $k^{th}$ decay mode. The dominant decay mode is ($t_R^X \tau_R$, $\overline{t_R^Y}$ $\overline{b_R^Z}$,  $\overline{t_R^Z}$ $\overline{b_R^Y}$)
while the most relevant mode for producing $e_R$'s is ($t_R^X e_R$, $\overline{t_R^Y}$ $\overline{d_R^Z}$, $\overline{t_R^Z}$ $\overline{ d_R^Y}$), X, Y, Z are color indices.
Hence, the number density of $e_R$ can at most be ${10}^{-6}$ (total
baryon/lepton number). For the baryon asymmetry produced via the sphaleron
conversion of $(e_R)$ lepton number to be the baryon asymmetry observed
today, the total baryon number produced in the heavy GUT scalar decay
should be $(n_B/s)_{\rm total} \sim {10}^{-5} - {10}^{-4}$, s is the
entropy density. Such a large value of $(n_B/s)_{\rm total}$ is just about
the maximum that is attainable through heavy scalar decay and requires a
large value of the CP-violation parameter $(\epsilon) \sim {10}^{-2}$
(possible only in the Weinberg Three-Higgs model [25]) and the GUT scalars
to be heavier than $({10}^{15} - {10}^{16})$ GeV so that they can decay
completely out of equilibrium. Nevertheless, survival of an adequately
large baryon asymmetry generated by the decay of heavy GUT scalars seems plausible.

For $m_{\nu_e} < {10}^{-3}$ eV, the out-of-equilibrium decay of $N_e$ can
produce a sufficiently large electron number [26] that can be converted
into the observed baryon number by sphaleron interactions. And if $M_e <
{10}^{10}$ GeV and $M_e <(M_{\mu}, M_{\tau})$, then the electron (baryon)
number does not face threat of being erased.

\vspace{.4cm}

{\bf 2.b.} In the presence of an $SU(2)_R \times SU(2)_L$ gauge symmetry C
and CP are not violated and, hence, a baryon asymmetry cannot be generated
[27]. Unification of gauge coupling constants is possible (in the absence
of split multiplets of fermions and supersymmetry) if the scale at which
$SU(2)_R$ breaks $({\Lambda}_R)$ is less than ${10}^{11}$ GeV [28]. So, the
decay of only the lightest GUT scalars (mass $\sim {10}^{10} - {10}^{11}$
GeV) present in the primeval soup can generate baryon asymmetry.

But when $SU(2)_R$ breaks, Majorana masses for the right-handed neutrinos
may be produced and lepton number violating scattering processes can be in
thermal equilbrium for $T \sim ({10}^{10} - {10}^{11})$ GeV. Further,
interactions mediated by $W_R$ are expected to be in thermal equilibrium
upto $T \sim ({10}^{-4} - {10}^{-2}) {\Lambda}_R ;{}^{\#5}$ and in the
presence of Majorana masses for the right-handed neutrinos they simply
equate the chemical potentials of all the charged right-handed leptons,
including $e_R$.

The combined effect of the rapid sphaleron interactions, the lepton number
violating processes and the interactions mediated by $W_R$ is to completely
erase any baryon (and lepton) asymmetry that may exist at $T \sim
({10}^{10} - {10}^{11})$ GeV.

If ${\Lambda}_R \ll ({10}^{10} - {10}^{11})$ GeV, then even the decay of
the lightest GUT scalars present in the primeval soup cannot produce a
baryon asymmetry. However, heavy GUT scalars can be produced around $T
{^<_{\! \sim}} {\Lambda}_R$ by the collapse or annihilation of topological
defects . But if $M_{\tau}$ (or even $M_{\mu}$ and $M_e$) is not much
smaller than ${\Lambda}_R$, its lepton number violating decay may be in
thermal equilibrium at the same time as the $W_R$ - mediated interactions
and, again, the baryon asymmetry produced in the vicinity of ${\Lambda}_R$
will be completely erased. Anyway, if the baryon asymmetry produced by the
monopoles and cosmic strings does, somehow, manage to survive to the
present as the observed asymmetry then ${\Lambda}_R >{10}^7$ GeV. This is
because the monopoles can annihilate efficiently only for $T {^>_{\!
\sim}} {10}^7$ GeV [29] and the annihilation of cusps on infinitely long
cosmic strings (this being the dominant mechanism) can generate an
adequately large baryon asymmetry only when $T> {10}^7 $ GeV [30].

As in the case without an $SU(2)_R$ gauge symmetry, the baryogenesis via
leptogenesis mechanism is also viable for the case with an $SU(2)_R$ gauge
symmetry. Only, now, ${\Lambda}_R$ should be larger than ${10}^6$ GeV
($M_{Z_R} > {10}^5$ GeV) to sufficiently suppress the destruction of $N_e$
due to interactions mediated by $W_R, Z_R$.${}^{\#6}$ But even here we do not
have any compelling reason for $m_{\nu_e} < {10}^{-3}$ eV, except that
this favors an out of equilibrium decay of $N_e$.$^{\#7}$

The lesson of the entire section 2 is that the baryon asymmetry generated
by the out-of-equilibrium decay of heavy GUT (type) scalars can only
barely survive to be the baryon asymmetry observed today, for
cosmologically interesting values of neutrino masses and mixing angles:
the likelihood of survival is really bleak in the presence of an $SU(2)_R$ gauge
symmetry at intermediate scales. The baryon asymmetry produced via the
out-of-equilbrium decay of $N_e$'s is more likely to survive, but this
requires $m_{\nu_e} <{10}^{-3}$ eV and we do not know if this inequality
really holds.

Lastly, the analysis of section 2 is meaningful only if EWPT is strongly
first order, and this requires an extended Higgs sector as for just one
Higgs doublet EWPT cannot be strongly first order if the vacuum is to be stable.

\vspace{.4cm}

{\bf 3.} Earlier, we had shown that working models can easily be
constructed that can protect the baryon asymmetry generated above $T_{EW}$
from the effect of sphaleron interactions and other lepton number
violating processes, and/or generate an adequately large baryon asymmetry,
below $T_{EW}$, well after the sphalerons have dropped out of thermal
equilibrium [14,15].

In one set of these models the heavy GUT scalars are constrained to decay
out of equilibrium during a temporary phase of broken electromagnetic
gauge invariance $U(1)_{em}$ [14]. The decay, thus, produces not just
non-zero values of $B$ and $L$ but also a non-zero electromagnetic charge
$Q$. Unless $Q$ and $(B-L)$ satisfy a specific relationship that depends
on the number of fermion generations and Higgs doublets, the sphalerons
cannot drive $B(L)$ to zero [15]. The electric charge neutrality of the
universe is restored, when $U(1)_{em}$ gauge invariance is restored
somewhere above $T_{EW}$, by the charge $-Q$ carried by the scalar field
$(\Phi)$ whose non-zero thermal expectation value is responsible for
breaking $U(1)_{em}$. The $\Phi$ stays out of chemical equilibrim till
well below $T_{EW}$ and then decays into leptons.

The other set of models [15] consists of simply modified versions of the
Weinberg Three-Higgs model [25]. Here, a temporary phase of broken
$U(1)_{em}$ is not required. An asymmetry in the numbers of two heavy GUT
scalars $\phi_1$ and $\phi_2$ is created by the CP-violating
out-of-equilibrium decays of a third heavy GUT scalar $\phi_3$. $\phi_1$
may have regular strength Yukawa couplings to the fermions and may decay
well above $T_{EW}$ to produce non-zero $B_1, L_1$ and $Q_1$, while
$\phi_2$ may only be weakly coupled to the fermions so that it remains out
of chemical equilibrium and can decay only after the sphalerons have
dropped out of thermal equilibrium well below $T_{EW}$. As before, the
non-zero electromagnetic charge $Q_1$ can 
protect the $B_1$ from sphalerons and lepton number violating processes.
The charge $Q_2(= -Q_1)$ carried by $\phi_2$, and eventually transferred
to the fermions it decays into, maintains the electric neutrality of the
universe throughout. The net baryon asymmetry is $B_1^{'} + B_2$, which is
generally non-zero and adequately large so that final $ n_B/s \sim (4-7)
\times {10}^{11}$ : $B_1^{'}$ is the sphaleron-modified value of $B_1$, while
$B_2$ is the baryon asymmetry produced in the $\phi_2 $ -decays.

Our models are fairly robust and capable of yielding an adequately large
baryon asymmetry, for any order of EWPT and any value of neutrino masses.

In section 2 we have noted that for a strongly first order EWPT, the
mechanism of EW baryogenesis can, at best, only marginally yield an
adequately large value of baryon asymmetry. While $B$ produced by the
conventional out-of-equilibrium decay of heavy GUT scalars at $T>T_{EW}$
just barely survives for cosmologically interesting values of neutrino
masses and mixing angles; the situation is particularly grim in the
presence of an $SU(2)_R$ gauge symmetry at intermediate scales. Only $B$
produced through the decay of right-handed electron neutrinos seems to
survive without any threat of destruction.

The exercise in equilibrium thermodynamics carried out in section 1 had
yielded our main result: for a second or weakly first order EWPT, $B(L)$
is made zero by the sphaleron interactions for non-zero values of Majorana
masses of neutrinos. This result holds for $B$ generated by any of the
well known mechanisms, including the baryogenesis via leptogenesis
mechanism. Only our models are still viable.

Since our models are variations on the old theme of {\it
out-of-equilibrium decays of heavy GUT (type) scalars can produce the
observed baryon asymmetry}, we are tempted, like Kolb and Turner [31], to
believe that the heavy GUT (type) scalars that link quarks and leptons,
though out of reach of experiments planned for the near future, may have
something to do with reality.

\pagebreak

\centerline{\large \bf FOOTNOTES}

\begin{description}

\item{\bf 1.} The Majorana mass eigenstates are actually $\omega (\chi)$
plus a small, $O(m_{d_i}/M_i)$, admixture of $\chi (\omega)$. We have
chosen to neglect this small admixture.

\item{\bf 2.} The largest value of CP-violation parameter $(\epsilon)$
attainable in the decay of heavy GUT gauge bosons is too small to yield
the observed baryon asymmetry [17].

\item{\bf 3.} The $2 \leftrightarrow 2 $ scatterings mediated by the
right-handed Majorana neutrinos will be in thermal equilibrium at a
temperature $T$ provided [13],

$$ T {^> _{\! \sim}}(4 eV/m_{\nu_i})^2 ({10}^{10} GeV).$$

\item{\bf 4.} For the decay to be in thermal equilibrium at $T \sim M$,
the decay rate $({\Gamma}_D)$ should be larger than the expansion rate of
the universe $H(T=M)$ :

$$\frac {1}{8 \pi}. \frac{m_d^2}{v^2}. M \;  {^>_{\! \sim}}\; 17. \frac
{M^2}{M_P} \ ;$$
using, $m_{\nu} = m_d^2/M$, $v=175 GeV$, we have $m_{\nu} {^>_{\! \sim}} 8
\times {10}^{-4}$ eV.

\item{\bf 5.} For $T< M_{W_R}$, the interactions mediated by $W_R$ are in
thermal equilibrium upto $T_R  {^>_{\!
\sim}}{({3.10}^3.{\Lambda}_R^4/M_P)}^{1/3}$. It should be noted that $T_R$
is independent of the $SU(2)_R$ gauge coupling constant, $g_R$, as
$M_{W_R} \sim g_R {\Lambda}_R$.
With ${\Lambda}_R = {10}^3$ GeV, $T_R \sim {10}^{-13/3} {\Lambda}_R$
and with ${\Lambda}_R = {10}^{11}$ GeV, $T_R \sim {10}^{-5/3} {\Lambda}_R$.

\item{\bf 6.} $N_e$ can annihilate via $Z_R$ mediated interactions and
the number that ultimately decay will be very small unless the decay rate
is larger than the annihilation rate at $T \sim M( < M_{Z_R})$, which
requires 
$${\Lambda}_R^4 > \frac {M^4}{2 {\pi}^2 f^2}$$
$f$ being the
electron Yukawa coupling $(\sim 2 \times {10}^{-6})$. If $M_i$ are to
decay before the sphalerons go out of equilibrium, then $M_i > {10}^2$ GeV
and hence $ {\Lambda}_R > ({10}^5 - {10}^6)$ GeV.

\item{\bf 7.} At present, the
double-beta decay experiments impose a limit: \linebreak $m_{\nu_e} < 0.68$ eV.

\end{description}

\pagebreak

\centerline{\large \bf REFERENCES }

\begin{description}

\item{[\bf 1]} see for example, D.O. Caldwell, Nucl. Phys. {\bf B} (Proc.
Suppl.) {\bf 43} (1995) 126; {\bf 48} (1996) 158.

\item{[\bf 2]} L. Wolfenstein, Phys. Rev. {\bf D 17} (1978) 2369;  \\
S.P. Mikheyev and A.Y. Smirnov, Sov. J. Nucl. Phys. {\bf 42} (1985) 913.

\item{[\bf 3]} Y. Fukuda et al., Phys. Lett. {\bf B 335} (1994) 237; \\
D.O. Caldwell, as in ref. [1].

\item{[\bf 4]} J.R. Primack et al., Phys. Rev. Lett. {\bf 74} (1995) 2160.

\item{[\bf 5]} M. Gell Mann, P. Ramond and R. Slansky, in Supergravity,
North Holland, Amsterdam, 1980.

\item{[\bf 6]} M. Fukugita and T. Yanagida, Phys. Rev. {\bf D 42} (1990)
1285; \\
A.E. Nelson and S.M. Barr, Phys. Lett. {\bf B 246} (1990) 141; \\
W. Fischler, G.F. Giudice, R.G. Leigh and S. Paban, Phys. Lett. {\bf B
258}(1991) 45; \\
J.A. Harvey and M.S. Turner, as in ref. [13].

\item{[\bf 7]} M. Shaposhnikov, Phys. Lett. {\bf B 277} (1992) 324; {\bf 282} (1992) 483 (E).

\item{[\bf 8]} K. Farakos, K. Kajantie, K. Rummukainen and M.
Shaposhnikov, Phys. Lett. {\bf B 336} (1994) 494.

\item{[\bf 9]} M. Sher, Phys. Lett. {\bf B 317} (1993) 159; {\bf 331}
(1994) 448 (E).

\item{[\bf 10]} M. Sher, Phys. Rep. {\bf 179} (1989) 274.

\item{[\bf 11]} A. Cohen, D. Kaplan and A. Nelson, Ann. Rev. Nucl. Part.
Phys. {\bf 43} (1993) 27.

\item{[\bf 12]} N. Turok, in Perspectives in Higgs Physics, edited by G.
Kane (World Scientific, Singapore, 1992), p. 300; \\
M. Yoshimura, preprint TU/96/500; hep-ph/9605246.

\item{[\bf 13]} J.A. Harvey and M.S. Turner, Phys. Rev. {\bf D42} (1990) 3344.

\item{[\bf 14]} S. Mohan, Phys. Lett. {\bf B 385} (1996) 169.

\item{[\bf 15]} S. Mohan, Phys. Lett. {\bf B} in press; hep-ph/9605329.

\item{[\bf 16]} H. Dreiner and G.G. Ross, Nucl. Phys. {\bf B 410} (1993) 188.

\item{[\bf 17]} S. Barr, G. Segre and H.A. Weldon, Phys. Rev. {\bf D 20}
(1979) 2494.

\item{[\bf 18]} M. Fukugita and T. Yanagida, Phys. Lett. {\bf B 174}
(1986) 45; \\
M. Luty, Phys. Rev. {\bf D 45} (1992) 455.

\item{[\bf 19]} J.M. Cline, K. Kainulainen and A.P. Vischer, Phys. Rev.
{\bf D 54} (1996) 2451; \\
M. Joyce, T. Prokopec and N. Turok, Phys. Rev. {\bf D 53} (1996) 2930; 2958.

\item{[\bf 20]} U. Amaldi, W. de Boer and H. Furstenau, Phys. Lett. {\bf B
260} (1991) 447.

\item{[\bf 21]} U. Amaldi, W. de Boer, P.H. Frampton, H. Furstenau and
J.T. Liu, Phys. Lett. {\bf B 281} (1992) 374.

\item{[\bf 22]} M.L. Kynshi and M.K. Parida, Phys. Rev {\bf D 49} (1994) 3711.

\item{[\bf 23]} Review of Particle Properties, Phys. Rev. {\bf D 54} (1996).

\item{[\bf 24]} J.M. Cline, K. Kainulainen and K.A. Olive, Phys. Rev.
Lett. {\bf 71} (1993) 2372; Phys. Rev. {\bf D 49} (1994) 6394; \\
K. Kainulainen, Nucl. Phys. {\bf B} (Proc. Suppl.) {\bf 43} (1995) 291.

\item{[\bf 25]} S. Weinberg, Phys. Rev. Lett. {\bf 37} (1976) 657.

\item{[\bf 26]} W. Fischler et al., as in ref. [6]; \\
M. Luty, as in ref. [18]; \\
M.Flanz, E.A. Paschos and U. Sarkar, Phys. Lett. {\bf B 345} (1995) 248.

\item{[\bf 27]} J.A. Harvey, E.W. Kolb, D.B. Reiss and S. Wolfram, Nucl.
Phys. {\bf B 210} (1982) 16; \\
E.W. Kolb and M.S. Turner, Ann. Rev. Nucl. Part. Sci. {\bf 33} (1983) 645.

\item{[\bf 28]} G. Amelino-Camelia, O. Pisanti and L. Rosa, Nucl. Phys.
{\bf B} (Proc. Suppl.) {\bf 43} (1995) 86.

\item{[\bf 29]} E. Gates, L.M. Krauss and J. Terning, Phys. Lett. {\bf B 284} (1992) 309; \\
S. Mohan, Mod. Phys. Lett. {\bf A 10} (1995) 227.

\item{[\bf 30]} M. Mohazzab, Phys. Lett. {\bf B 350} (1995) 13.

\item{[\bf 31]} E.W. Kolb and M.S. Turner, The Early Universe
(Addison-Wesley, Redwood City, CA, 1990), Ch. 6, last sentence.

\end{description}

\end{document}